\documentclass[aps,prb,twocolumn,superscriptaddress,showpacs]{revtex4}
\usepackage{graphicx}

\begin{document}

\title{Superconductivity of Mo$_{3}$Sb$_7$ from first principles}

\author{B. Wiendlocha}
\email[Corresponding author: ]{bartekw@fatcat.ftj.agh.edu.pl}
\affiliation{Faculty of Physics and Applied Computer Science, AGH
University of Science and Technology, Al. Mickiewicza 30, PL-30059
Cracow, Poland}

\author{J. Tobola}
\affiliation{Faculty of Physics and Applied Computer Science, AGH
University of Science and Technology, Al. Mickiewicza 30, PL-30059
Cracow, Poland}

\author{M. Sternik}
\affiliation{Institute of Nuclear Physics, Polish Academy of
Sciences, Radzikowskiego 152, PL-31342 Cracow, Poland}

\author{S. Kaprzyk}
\affiliation{Faculty of Physics and Applied Computer Science, AGH
University of Science and Technology, Al. Mickiewicza 30, PL-30059
Cracow, Poland}

\author{K. Parlinski}
\affiliation{Institute of Nuclear Physics, Polish Academy of
Sciences, Radzikowskiego 152, PL-31342 Cracow, Poland}

\author{A. M. Ole\'s}
\affiliation{Institute of Nuclear Physics, Polish Academy of
Sciences, Radzikowskiego 152, PL-31342 Cracow, Poland}

\date{\today}

\begin{abstract}
Superconductivity in Mo$_3$Sb$_7$ is analyzed using the combined
electronic structure and phonon calculations, and the {\em
electron--phonon coupling} constant $\lambda_{\rm ph}=0.54$ is
determined from first principles. This value explains the
experimental value of the superconducting critical temperature
$T_c=2.2$~K. The possible influence of spin fluctuations and spin
gap on the superconductivity in Mo$_3$Sb$_7$ is discussed, and
electron--paramagnon interaction is found to be weak.
\end{abstract}



\maketitle

A paramagnetic intermetallic compound Mo$_3$Sb$_7$ is a type II
superconductor,\cite{Buk02,Buk06} with the critical temperature
$T_c\simeq 2.2$~K. The temperature characteristics of the specific
heat, the superconducting gap, and the magnetic critical field
suggest that the conventional electron--phonon interaction might
be responsible for the superconductivity.
\cite{cc-prl,Buk06,tran-heat,cc-heat} Recently, however, Candolfi
{\it et al.}\cite{cc-prl} argued that spin fluctuations (SFs) are
present in Mo$_3$Sb$_7$. This interpretation is supported by two
unusual features: (i) the quadratic temperature dependence of both
electrical resistivity and magnetic susceptibility, as well as
(ii) the high value of the susceptibility at room temperature.
They also reported a much smaller value of the electronic specific
heat jump\cite{cc-prl} at the transition point $\Delta C/\gamma
T_c = 1.04$ than the weak--coupling BCS value 1.43, which might
suggest additional enhancement of the electronic specific heat
coefficient by the SFs. Very recently, Tran {\it et al.}
\cite{Tra08} observed a peak in the specific heat $C_P(T)$ at
$T^*=50$~K, which was interpreted as supporting the presence of
spin gap. Also, they explained the anomalous behavior of the
magnetization and resistivity in terms of the gap opening.
Moreover, they analyzed the electronic specific heat in the
superconducting state in terms of the two BCS gap model,
\cite{tran-heat} and reported a higher value of $\Delta C/\gamma
T_c=1.56$ than the one measured before.\cite{cc-prl}

In order to elucidate the possible origin of superconductivity, an
{\em ab initio} approach which involves the electronic structure
and phonon calculations may be used to determine the 
electron--phonon coupling (EPC) constant. For instance, a recent
determination of the EPC constant suggested that the
superconductivity in PuCoGa$_5$ is driven by an unconventional
mechanism based on antiferromagnetic (AF) fluctuations.\cite{Pie05}
Here we present an {\em ab initio} study of the EPC constant and
superconductivity in Mo$_3$Sb$_7$, where SFs might play a role.
The electron--phonon interaction is treated within the rigid
muffin tin (MT) approximation. The superconducting critical
temperature $T_c$ and its possible modification by SFs is
discussed using two approaches: (i) the McMillan
formula,\cite{mcm,allen} and (ii) the equation for $T_c$ which
explicitly includes the presence of paramagnons.\cite{Dol05}

Electronic structure calculations were performed using the
Korringa-Kohn-Rostoker (KKR) multiple scattering
method.\cite{kkr99} The crystal potential was constructed in the
framework of the local density approximation (LDA), using von
Barth and Hedin formula \cite{lda} for the exchange--correlation
part. For all atoms angular momentum cut--off $l_{max}=4$ was set;
\makebox{{\bf k}--point} mesh in the irreducible part of the
Brillouin zone (BZ) contained about 400 points. Density of states
(DOS) was computed using the tetrahedron \makebox{{\bf k}--space}
integration technique, generating about 1500 tetrahedrons in the
irreducible part of the BZ. Semirelativistic calculations results
are presented here. Since our main goal in this work is to
estimate the EPC constant from first principles within the rigid
MT approximation, spherical potential approximation for the
crystal potential is used, as is required in this approach.
Mo$_3$Sb$_7$ crystallizes in a cubic {\it bcc} structure (space
group {\it Im3m}) of the Ir$_3$Ge$_7$ type, with lattice
constant\cite{cc-cryst} $a=9.58$ \AA. The primitive cell of
Mo$_3$Sb$_7$ contains two formula units, i.e. 20 atoms, occupying
three nonequivalent positions: Mo in (12e) with $x=0.3432$, Sb(1)
in (12d) and Sb(2) in (16f) with $x=0.1624$.

The phonon frequencies were determined within the direct
method,\cite{Par97} which utilizes Hellmann--Feynman forces
obtained by performing small atomic displacements of nonequivalent
atoms from their equilibrium positions. From them the dynamical
matrix is determined and diagonalized to obtain the phonon
frequencies at each wave vector. The crystal structure
optimization and calculations of the complete set of
Hellmann--Feynmann forces were performed using the
first-principles {\sc vasp} package\cite{vasp} which makes use of
the Perdew, Burke, and Ernzerhof (PBE) functional.\cite{Per96} The
calculations were performed on a $\sqrt{2}\times\sqrt{2}\times 1$
supercell (containing 80 atoms) 
with periodic boundary conditions. The wave functions
were sampled according to Monkhorst--Pack scheme with a {\bf
k}--point mesh of (4,4,4). After the optimization we obtained the
lattice parameter $a=9.6405$~\AA~and the atomic positions of
(0.3421,0,0), (0.25,0,0.5) and (0.1608,0.1608,0.1608) for Mo,
Sb(1) and Sb(2), respectively. The determined values are in very
good agreement with the experimental data.\cite{cc-cryst}


\begin{figure}[b!]
\includegraphics[width=8.0cm]{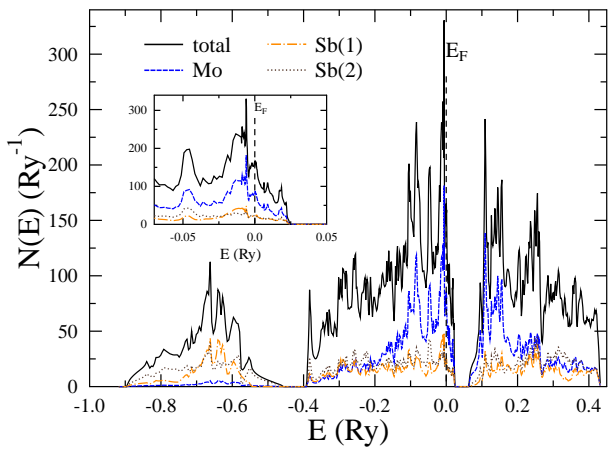}
\caption{\label{fig-tdos}(Color online) Total and site-decomposed
densities of electronic states in Mo$_3$Sb$_7$ (per formula unit).
The inset shows the details of the DOS near $E_F=0$.}
\end{figure}

The electronic structure was computed for the experimental lattice 
parameters and atomic positions. Total and site--decomposed electronic 
DOSs of Mo$_3$Sb$_7$ are presented in Fig.~\ref{fig-tdos}. The most 
intriguing feature of the electronic spectrum is the presence of a 
narrow band gap just above the Fermi level, with $E_F$ located in 
the range of sharply decreasing DOS.\cite{noteef}
In the inset in Fig.~\ref{fig-tdos} one observes that $E_F$
coincides with a local DOS maximum. By analyzing the angular
contributions to the total DOS at $E_F$, presented in
Table~\ref{tab:1}, we deduced that the bands near $E_F$ are built
out of the Mo$(4d)$ and Sb$(5p)$ states. The largest atomic
contribution comes from Mo atom, with the value $n_{\rm
Mo}(E_F)\simeq 14$~Ry$^{-1}$/spin, being not far but below the
magnetic instability [the computed Stoner parameter $I$ satisfies
$In_{\rm Mo}(E_F)\simeq 0.7$]. Note that the spin--polarized KKR
calculations assuming ferromagnetic (FM) spin order led to the
nonmagnetic ground state.

Tran {\it et al.}\cite{Tra08} suggested the opening of the spin
gap below 50~K, caused by the AF interactions between the selected
nearest pairs of Mo atoms. They argued that these atoms form
dimers, and the AF interaction stabilizes there spin singlets (but
long--range order is absent). We examined a few possible AF
structures for this compound, e.g. with alternating moments in Mo
planes, but stable AF configuration could not be reached and all
magnetic moments converged to zero values. Note, that the proposed
model,\cite{Tra08} including one AF and two FM types of Mo--Mo
interactions, creates a geometrical frustration of the Mo
sublattice. The high value of the DOS at $E_F$, as well as the
suggested different magnetic interactions between Mo atoms, may
also give rise to the SFs, which could appear in real sample.

The electronic structure results were used to calculate the
electronic part of the EPC constant, i.e. the McMillan--Hopfield
$\eta_i$ parameters\cite{mcm,hop} for each atom. They follow from
the formula:\cite{rmta,pickett}
\begin{equation}
\label{eq:eta} \eta_i =\!\sum_l \frac{(2l + 2)\,n_l\,
n_{l+1}}{(2l+1)(2l+3)N(E_F)} \left|\int_0^{R_{\mathsf{MT}}}\!\!r^2
R_l\frac{dV}{dr}R_{l+1} \right|^2\!,
\end{equation}
where $V(r)$ is the self--consistent potential at site $i$,
$R_\mathsf{MT}$ is the radius of the $i$-th MT sphere, $R_l(r)$ is
a regular solution of the radial Schr\"odinger equation
(normalized to unity inside the MT sphere), $n_l(E_F)$ is the
$l$--th partial DOS per spin at the Fermi level $E_F$, and
$N(E_F)$ is the total DOS per cell and spin. The values of
$\eta_i$ parameters (\ref{eq:eta}), with contributions from each
$l\rightarrow l+1$ scattering channels, are presented in
Table~\ref{tab:1}. For Mo, the $d$--$f$ channel is the most
important one (typically for $d$--element), whereas $p$--$d$
contribution dominates for both Sb atoms. The Sb(1) and Sb(2)
atoms have very similar $\eta_i$ parameters, despite quite
different $p$--DOSs. This is a result of opposite behavior in both
partial DOSs, i.e. for the Sb(2) atom the lower $p$--DOS is
compensated by the larger value of $d$--DOS [the radial wave
functions matrix elements form Eq.~(\ref{eq:eta}) are similar in
both cases].

\begin{figure}[b!]
\includegraphics[width=7.7cm]{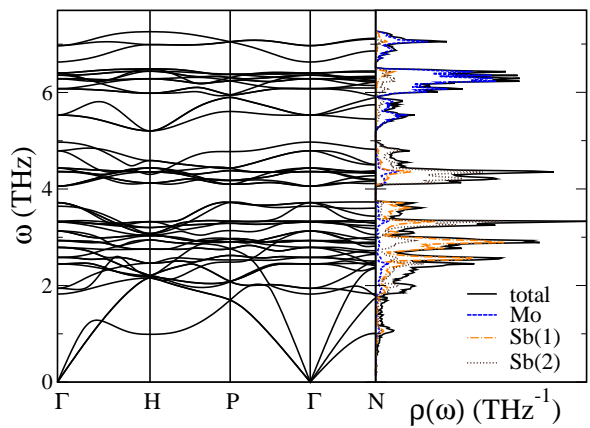}
\caption{\label{fig-phdos}(Color online) Phonon dispersions along
the high symmetry directions of the BZ (left) and total and
site--decomposed densities of phonon states in Mo$_3$Sb$_7$
(right). The special points are: $\Gamma=(0,0,0)$,
$H=(1/2,-1/2,1/2)$, $P=(3/4,-1/4,3/4)$, $\Gamma=(1,0,1)$,
$N=(1,0,1/2)$.}
\end{figure}

\begin{table*}[t!]
\caption{Site--decomposed electronic and dynamic properties of
Mo$_3$Sb$_7$. $n_i(E_F)$ is in Ry$^{-1}$/spin, $\eta_i$ in
mRy/${a_B}^2$ (both per atom), $\omega_i$ in THz. Values of
$\lambda_i$ take into account the number of $i$--type atoms in the
primitive cell: 6 Mo, 6 Sb(1), 8 Sb(2).} \label{tab:1}
\begin{ruledtabular}
\begin{tabular}{lccccccccccc}
atom & $n_i(E_F)$ & $n_s(E_F)$ &$n_p(E_F)$ &$n_d(E_F)$ &
$n_f(E_F)$ & $\eta_i$ &
$\eta_{sp}$ & $\eta_{pd}$ & $\eta_{df}$ & $\sqrt{\langle\omega_i^2\rangle}$  & $\lambda_i$\\
\hline
Mo    & 14.3 & 0.05& 0.54 & 13.7 & 0.04  & 6.75 &  0.0 & 1.3 & 5.4 & 5.07 & 0.19  \\
Sb(1) &  3.7 & 0.09& 3.16 & 0.37 & 0.06  & 2.64 &  0.0 & 2.6 & 0.0 & 2.97 & 0.17  \\
Sb(2) &  3.0 & 0.13& 2.19 & 0.51 & 0.17  & 2.60 &  0.0 & 2.6 & 0.0 & 3.41 & 0.17  \\
\end{tabular}
\end{ruledtabular}
\end{table*}

The phonon dispersion relations along the high symmetry
directions and the total and site--decomposed partial phonon DOSs
were computed by random sampling of the BZ and are presented in
Fig.~\ref{fig-phdos} for the optimized supercell. The optic
phonons give three characteristic maxima of the phonon DOS
$\rho(\omega)$ at $\omega\simeq 2.8$, 4.4, and 6.3 THz. The Mo
atoms, which are about 30\% lighter than Sb atoms, contribute
mainly to the high frequency part of the phonon DOS. The phonon
DOS was used to compute the average square site--decomposed phonon
frequencies $\langle\omega_i^2\rangle$ presented as well in
Table~\ref{tab:1}. These quantities, together with $\{\eta_i\}$
parameters, are needed to deduce the EPC constant
\begin{equation}
\label{eq:lam} \lambda_{\rm ph} = \sum_i \frac{\eta_i}{M_i\langle
\omega_i^2 \rangle} = \sum_i \lambda_i.
\end{equation}
Here $i$ runs over all the atoms in the primitive cell and $M_i$
is the atomic mass. For a review, more detailed discussion of the 
approximations involved in this approach, and a number of references 
to the previous rigid MT studies, see e.g. Ref.~\onlinecite{prb-bw} 
and references therein.

Surprisingly, one finds that all the atoms are equally important
for the onset of superconductivity in Mo$_3$Sb$_7$. The
contribution from Mo atoms to the total $\lambda_{\rm ph}$,
despite the dominant character of Mo states near $E_F$, is only
slightly larger than those from Sb(1) and Sb(2) respectively. This
is a consequence of higher partial phonon frequencies for Mo. It
is worth noting that Sb(1) and Sb(2) have the same $\lambda_i$
values in spite of rather different average phonon frequencies.
Here, the effect of higher ${\langle\omega_i^2\rangle}$ for Sb(2)
is compensated by the larger multiplicity of this crystallographic
site. The calculated total EPC constant (\ref{eq:lam}) is
$\lambda_{\rm ph}=0.54$, which qualifies Mo$_3$Sb$_7$ as the
medium--coupling superconductor.

We estimated the superconducting critical temperature $T_c$ using
two formulas: (i) a McMillan--type formula,\cite{mcm,allen} with
the logarithmically averaged phonon frequency $\omega_{\rm
ph}\equiv\langle\omega_{\rm log}\rangle$ in the prefactor,
\begin{equation}\label{eq:tc}
T_c =  \frac{\omega_{\rm ph}}{1.20}\,\exp\left\{
-\frac{1.04(1+\lambda\mathsf{_{eff}})}
{\lambda\mathsf{_{eff}}-\mu\mathsf{_{eff}}^{\star}
(1+0.62\lambda\mathsf{_{eff}})}\right\},
\end{equation}
and (ii) the formula including the interaction of electrons with
paramagnons, and successfully applied before to
MgCNi$_3$,\cite{Dol05}
\begin{eqnarray}\label{eq:tc2}
T_c\!&=&\!1.14\;\omega_{\rm ph}^{\lambda_{\rm ph}/(\lambda_{\rm
ph}-\lambda_{\rm sf})}\;
\omega_{\rm sf}^{-\lambda_{\rm sf}/(\lambda_{\rm ph}
-\lambda_{\rm sf})}\;e^K \nonumber \\
\!& \times &\!\exp \left\{ - \frac{1+\lambda_{\rm ph} +
\lambda_{\rm sf}} {\lambda_{\rm ph} - \lambda_{\rm sf} -
\mu^{\star} (1-K\frac{\lambda_{\rm ph}-\lambda_{\rm
sf}}{1+\lambda_{\rm ph}+\lambda_{\rm sf}})}\right\},
\\
K\!&=&\! - \frac{1}{2} - \frac{\lambda_{\rm ph}\lambda_{\rm sf}}
{(\lambda_{\rm ph} - \lambda_{\rm sf})^2}\left[1+\frac{\omega_{\rm
ph}^2+\omega_{\rm sf}^2} {\omega_{\rm ph}^2-\omega_{\rm
sf}^2}\ln\frac{\omega_{\rm sf}}{\omega_{\rm ph}}\right].
\end{eqnarray}
Here $\lambda_{\rm sf}$ stands for the electron--paramagnon
interaction parameter, and $\omega_{\rm sf}$ is the characteristic
SF frequency (temperature).

The interplay between SFs and superconductivity is a well--known
problem in the theory of superconductivity. In conventional
superconductors, with electron--phonon pairing mechanism, FM SFs
are known to compete with superconductivity, leading e.g. to the
lack of superconductivity in palladium.\cite{berk} More recently,
SFs (paramagnons) were studied in the context of superconductivity
in MgCNi$_3$,\cite{Dol05} or for elemental metals under pressure:
Fe,\cite{Maz02,Jar02} and Sc.\cite{Bos08} In fact, one finds that
in case of SF superconductor the McMillan formula\cite{mcm,allen}
may still be used, but the parameters $\lambda$ and $\mu^{\star}$,
applied when $\lambda_{\rm sf}=0$ in Eq. (\ref{eq:tc}), are then
renormalized to:\cite{Car81} $\lambda\mathsf{_{eff}} =
\lambda_{\rm ph}/(1+\lambda_{\rm sf})$,
$\mu\mathsf{_{eff}}^{\star} = (\mu^{\star} + \lambda_{\rm
sf})/(1+\lambda_{\rm sf}).$

\begin{figure}[b!]
\includegraphics[width=7.5cm]{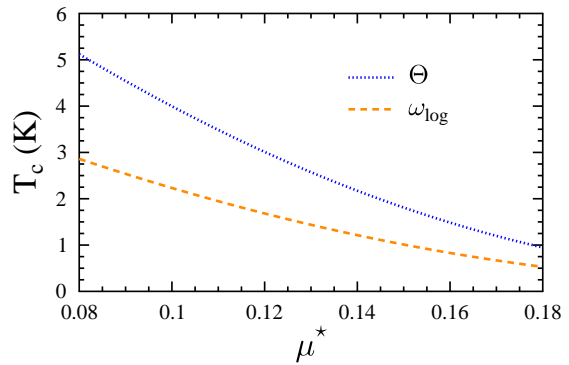}
\caption{\label{fig-tc-mu}(Color online) Critical temperature
$T_c$ as a function of Coulomb parameter $\mu^{\star}$ for
$\lambda_{\rm sf}=0$ and two prefactors in Eq. (\ref{eq:tc}):
$\omega_{\rm ph}/1.20$, and $\Theta/1.45$. Parameters:
$\lambda_{\rm ph}=0.54$, $\lambda_{\rm sf}=0$, $\omega_{\rm
ph}=143$ K, $\Theta=310$~K.}
\end{figure}

First we calculate $T_c$ without taking into account the SFs, i.e.
using Eq.~(\ref{eq:tc}) with $\lambda\mathsf{_{eff}}=\lambda_{\rm
ph}$ and $\mu\mathsf{_{eff}}^{\star}=\mu^{\star}$. Since the value
of Coulomb pseudopotential parameter $\mu^{\star}$ is unknown, we
present $T_c$ in a realistic range of $0.08<\mu^{\star}<0.18$ in
Fig.~\ref{fig-tc-mu}. For the typical values of $\mu^{\star}$ and
the calculated $\omega_{\rm ph}=143$~K we get $T_c=2.4$~K
($\mu^{\star}=0.10$) and 1.6~K ($\mu^{\star}=0.13$). Note, that
when the prefactor in the McMillan equation is set to the original
value\cite{mcm} $\Theta/1.45$, and the experimental
value\cite{cc-prl} of Debye temperature $\Theta=310$~K is used,
the resulting temperatures are higher: $T_c = 4.0$~K
($\mu^{\star}$ = 0.10), 2.6~K ($\mu^{\star}$ = 0.13), 1.8~K
($\mu^{\star}$ = 0.15). These results demonstrate that, depending
on the prefactor, the experimental critical temperature $T_c=2.2$
K may be explained using the EPC constant $\lambda_{\rm ph}=0.54$
derived within the rigid MT approximation, and taking
$\mu^{\star}$ between 0.10 and 0.13.

\begin{figure}[t!]
\includegraphics[width=7.5cm]{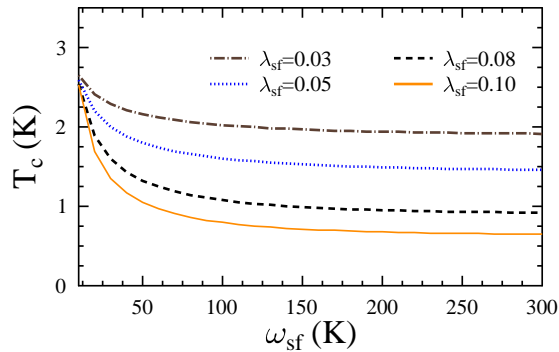}
\caption{\label{fig-tc}(Color online) Critical temperature $T_c$
as obtained from Eq. (\ref{eq:tc2}) for increasing paramagnon
frequency $\omega_{\rm sf}$ and for different values of
$\lambda_{\rm sf}$. Parameters: $\lambda_{\rm ph}=0.54$,
$\mu^{\star}=0.09$, $\omega_{\rm ph}=143$ K.}
\end{figure}

Next we analyze the possible influence of SFs on the transition
temperature $T_c$. The electron--paramagnon mass enhancement
$\lambda_{\rm sf}$ is treated as a parameter. It is important to
note that if one explicitly takes into account the SF effect on
the superconductivity, the starting value of $\mu^{\star}$ (i.e.
before its renormalization by $\lambda_{\rm sf}$) can be taken
smaller than typically used (e.g. for Nb $\mu^{\star}=0.086$ was
used in Ref.~\onlinecite{Car81}). Since Eq.~(\ref{eq:tc2})
involves additional parameter, i.e. the characteristic paramagnon
frequency $\omega_{\rm sf}$, in this case we plotted $T_c$ against
$\omega_{\rm sf}$ for some representative values of $\lambda_{\rm
sf}$ in Fig.~\ref{fig-tc}. For $\omega_{\rm sf}>100$~K the value
of $T_c$ is practically nonsensitive to the chosen $\omega_{\rm
sf}$, thus this value was used in the calculations. Using
Eq.~(\ref{eq:tc2}), one finds that temperatures close to the
observed $T_c=2.2$~K may be obtained for $\lambda_{\rm sf}=0.03$
and $\mu^{\star}=0.08 - 0.09$ (corresponding to the effective
$\mu^{\star}_{\mathsf{eff}}=0.11-0.12$), i.e. $T_c=2.3$~K and
$T_c=2.0$~K, respectively. Fig.~\ref{fig-tc} shows that $T_c$
quickly tends below 2~K when the electron--paramagnon interaction
parameter $\lambda_{\rm sf}\geq 0.05$.\cite{notemu} Thus we conclude
that the observed magnitude of the superconducting critical
temperature can be explained taking into account the SF effects,
but the $\lambda_{\rm sf}$ parameter has to be relatively small,
$\lambda_{\rm sf}\simeq 0.03$, if the EPC parameter $\lambda_{\rm
ph}=0.54$ obtained in our study is used.

Another interesting question concerns the influence of the spin
gap, detected below $T^*=50$~K, on the superconducting state of
Mo$_3$Sb$_7$. In view of the present results, this effect cannot
be very strong, since (i) not all the Mo atoms are involved in
building the singlet dimers (responsible for the gap\cite{Tra08}),
and (ii) the Mo sublattice contribution to the total EPC constant
$\lambda_{\rm ph}$ is about 35\%, with the rest provided by the
two Sb sublattices.

In summary, the results of electronic structure and phonon
calculations were used to calculate the $\lambda_{\rm ph}$
parameter for the spin--fluctuation/spin--gap superconductor
Mo$_3$Sb$_7$, within the rigid MT approximation. The estimated
value of $\lambda_{\rm ph}=0.54$ qualifies Mo$_3$Sb$_7$ as a
medium--coupling superconductor. The experimentally observed
critical temperature $T_c\simeq 2.2$~K may be correctly reproduced
even including the presence of paramagnons, with small
$\lambda_{\rm sf}\simeq 0.03$. Thus, the spin fluctuations may
exist in Mo$_3$Sb$_7$, but the electron--paramagnon interaction
has to be moderate. Since the Mo contribution to the constant
$\lambda_{\rm ph}$ is estimated to be comparable to Sb(1) and
Sb(2) sublattices, the possible influence of spin gap on the
superconductivity is expected to be rather weak. However, in the
range of the EPC constant $\lambda_{\rm ph}\sim 0.5$ the value of
$T_c$ is quite sensitive even to small changes in $\lambda_{\rm
ph}$, so a more quantitative explanation of the superconductivity
in Mo$_3$Sb$_7$ requires further study.

This work was partly supported by the Polish Ministry of Science
and Education under Projects No. 44/N-COST/2007/0 and N202 1975 33. A.M.O
acknowledges support by the Foundation for Polish Science (FNP).


\begin{thebibliography}{99}
\bibitem{noteef} The position of $E_F$ close to such a sharp peak makes
                the $N(E_F)$ value sensitive to the computational details.
                In present calculations, using spherical potential
                approximation, gave $N(E_F)\simeq 165$ Ry$^{-1}$,
                whereas our full potential KKR calculations resulted in
                $N(E_F)\simeq 141$ Ry$^{-1}$.
\bibitem{notemu} If we use instead renormalized Eq.~(\ref{eq:tc}) we get
                 $T_c=2.0$~K for $\mu^{\star}_{\mathsf{eff}}=0.11$ and
                 $T_c=1.7$~K for $\mu^{\star}_{\mathsf{eff}}=0.12$
                 respectively.
\end{thebibliography}
\end{document}